\documentclass[sigconf]{acmart}

\AtBeginDocument{%
  \providecommand\BibTeX{{%
    \normalfont B\kern-0.5em{\scshape i\kern-0.25em b}\kern-0.8em\TeX}}}

\setcopyright{acmcopyright}
\copyrightyear{2018}
\acmYear{2018}
\acmDOI{XXXXXXX.XXXXXXX}

\acmConference[Conference acronym 'XX]{Make sure to enter the correct
  conference title from your rights confirmation email}{June 03--05,
  2018}{Woodstock, NY}
\acmPrice{15.00}
\acmISBN{978-1-4503-XXXX-X/18/06}

\usepackage{dcolumn}
\newcolumntype{d}[1]{D{.}{.}{#1}}

\copyrightyear{2022}
\acmYear{2022}
\setcopyright{acmlicensed}\acmConference[MUM 2022]{21th International Conference 
on Mobile and Ubiquitous Multimedia}{November 27--30, 2022}{Lisbon, Portugal}
\acmBooktitle{21th International Conference on Mobile and Ubiquitous Multimedia 
(MUM 2022), November 27--30, 2022, Lisbon, Portugal}
\acmPrice{15.00}
\acmDOI{10.1145/3568444.3568457}
\acmISBN{978-1-4503-9820-6/22/11}

\begin{document}

\title{Obtrusive Subtleness and Why We Should Focus on Meaning, not Form, in Social Acceptability Studies}


\author{Alarith Uhde}
\email{alarith.uhde@uni-siegen.de}
\orcid{0000-0003-3877-5453}
\affiliation{%
  \institution{University of Siegen}
  \streetaddress{Kohlbettstraße 15}
  \city{Siegen}
  \country{Germany}
  \postcode{57072}
}

\author{Tim zum Hoff}
\email{tim.zumhoff@uni-siegen.de}
\orcid{0000-0002-1785-8228}
\affiliation{%
  \institution{University of Siegen}
  \streetaddress{Kohlbettstraße 15}
  \city{Siegen}
  \country{Germany}
  \postcode{57072}
}

\author{Marc Hassenzahl}
\email{marc.hassenzahl@uni-siegen.de}
\orcid{0000-0001-9798-1762}
\affiliation{%
  \institution{University of Siegen}
  \streetaddress{Kohlbettstraße 15}
  \city{Siegen}
  \country{Germany}
  \postcode{57072}
}


\begin{abstract}

Nowadays, interactive technologies are used almost everywhere. As a result,
  designers need to increasingly make them ``socially acceptable''. Previous
  work recommends ``subtle'' forms of interaction to increase social
  acceptability and avoid negative experiences. Although often appropriate, such
  uniform recommendations neglect the variety of social situations. We
  demonstrate this limitation in an experiment (N=35), by comparing the observer
  experience of different forms of interaction in ``face-to-face
  conversations'', a social situation rarely studied. Here, the typically
  recommended form of interaction (``subtle'') led to a more negative observer
  experience than the usually deprecated form (``suspenseful''), in terms of
  affective experience and product perception. It also made the user appear less
  extraverted. We conclude by positioning interactions with technology not as
  separate from the social situation in which they are performed, but as
  a constitutive part of it that meaningfully relates to other situated
  activities.

\end{abstract}



\maketitle

\section{Introduction}

Technology is often used in social situations. For example, people interact with
their smartphones in public, and it is often inevitable that others observe or
witness such interactions in some way. The presence of others adds the challenge
for designers to make interacting with technology ``socially acceptable''. This
is particularly difficult, because the specific social situations we encounter
in everyday life vary considerably: People, their activities, social and
cultural norms, and the technology itself, all come together to form a wide
variety of potential social situations. Often, users are intuitively aware of
the differences between social situations and adapt their interactions with
technology accordingly, if possible. They know, for example, that speaking
loudly on the phone is not appreciated in a library, but quite acceptable on
a busy shopping street. Given the complex and nuanced differences between social
situations, it seems evident that depending on the particularities of each,
different forms of interaction may have different impact on people's
experiences.

However, current recommendations in Human-Computer Interaction (HCI) for
socially acceptable interactions are quite uniform. Social acceptability
guidelines mostly recommend inconspicuous or hidden forms of interaction to make
them as unobtrusive as possible. For example, Koelle and
colleagues~\citep{Koelle2020} found that the most common strategy to design for
social acceptability is to make interactions ``subtle'', that is, unobtrusive,
not drawing attention, and possibly disguised as everyday
activities~\citep{Pohl2019, Rico2010}. In some cases, the interaction can even
be designed to be entirely invisible to ``observers'' (i.e., from other people's
point of view), as a way to increase social acceptability. Yet another approach
is to hide the interactive devices in accessories or jewelry \citep{Ogata2012,
Rekimoto2001}. Of course, such unobtrusive and subtle forms of interaction seem
compelling, because they can be used across many social situations. At best,
they are not even registered by others, and thus do not offend anyone (i.e.,
neither in the library nor the shopping street). But this also leads to
a general design trend to remove the interaction from social situations, rather
than explicitly addressing and catering for them. In other words, instead of
``designing for the social'', current social acceptability guidelines to a large
extend follow a strategy to ``design despite the social''.

Although the current recommendation for subtle, unobtrusive, or even hidden
forms of interaction may at first seem like a safe choice, we believe that it
comes with its own problems. First, by removing the interaction from the social
space, we abandon its potential to create or promote positive social
experiences, or to contribute to positive social change. Second, the variety of
social situations we encounter in everyday life can be overwhelming. Designers
might choose to take such standard recommendations as a shortcut to avoid
engaging with this variety, and resort to boilerplate solutions, assuming that
unobtrusiveness will at least create no harm. But in fact, there may be social
situations where the seemingly unobtrusive forms of interaction are actually
experienced negatively. It simply seems implausible that the same forms are
appropriate across all social situations and activities, from libraries to rock
concerts, and from family dinners to face-to-face conversations with friends
about hobbies. After all, ``non-technical'' forms of activities in these
different situations also vary considerably, and it is unclear why the rules for
interactions with technology should be different.

Of course, the current recommendation for subtleness is based on empirical
findings, from social situations where subtleness consistently turned out to be
a good design strategy. Thus, the main contribution of our paper is an
experimental study from a different, specifically selected social situation,
where subtleness may not be the best choice. In fact, in our study a subtle
interaction led to a more negative experience, compared to an alternative,
open and outgoing form of interaction, which is described as especially
problematic in the literature (i.e., a ``suspenseful'' interaction, explained in
detail below). This contradicts the current recommendations for subtle and
against these ``suspenseful'' and more expressive forms of interaction. The
second contribution is a critical reflection on the apparent empirical
contradiction between our findings and previous work. Generally speaking, we
think that focusing more on the situated meaning of forms of interaction, and
less on the forms themselves, could move the field forward and allow us to more
specifically cater for a broader range of social situations.

In the following, we first review the crucial literature concerned with social
situations and how they relate to technology-mediated experiences. We then
present our experimental study in which we compared the effects of different
forms of casual interaction on the non-interacting partners (``observers'')
during a face-to-face conversation.  Finally, we discuss our findings in light
of the previous work and sketch a conceptual model with a focus on co-located,
meaningfully interrelated activities.

\section{Background}

\subsection{Understanding Social Situations}


In the late 1950s, Erving Goffman developed still highly influential theoretical
work about social situations~\citep{Goffman1959}. He described people as
``performers'', who attempt to leave a positive impression on others, that is,
their ``audience''. Goffman looked at how people strategically control their
behavior in public and private situations. He argued that in public situations
(e.g., a restaurant or a park), people perform for each other to leave positive
impressions. In contrast, private situations (e.g., a home) allow them to
perform less flattering activities, and to prepare their public performances. He
also described a few cases where the public and the private overlap, for example
when having guests at home. Nowadays, due to current communication technologies,
overlaps of the public and the private have become the norm. For example, people
use social media in their bedroom or stream videos from their living room,
making these situations ``public in private''. Conversely, texting makes private
conversations possible even in public, surrounded by strangers~\citep{Loh2020}.
Thus, the public and the private are increasingly difficult to tell apart.

Given the sheer variety of potential public and, thus, social situations, it is
difficult to identify the situational aspects relevant for design. In HCI, two
broad approaches to understanding the relevant situational factors can be
distinguished~\citep{Dourish2004}. The first one understands social contexts or
situations as a ``representational problem''. Its proposed solution is to
capture all the details of a social situation as separate pieces of information
in the sense of attributes. In this view, a social situation can be represented
as ``a bar with 30 people, dim light, background music, and table service''.
Technology is then supposed to make use of this information and, for example, to
adapt itself to the situation (e.g., through
``context-awareness''~\citep{Dey2001, Schmidt1999}).

The other approach understands situation as an ``interactional problem''. In
this line of thinking, a social situation is marked by the relations between the
activities performed by all the people involved. In contrast to the
representational approach, interactions with technology do not happen ``within''
a social situation, but the activities themselves constitute the situation,
thereby becoming an inseparable part of it. For example, imagine the same bar as
above. People sit, chat, and drink. At some point, someone turns the music
louder and a few people start to dance, others join in. Their collective
behavior transforms the bar into something like a club. Initially, everything
else remains unchanged, except for the volume of the music. But this new
attribute alone (i.e., ``loud music'') does not change much. It only renders
some activities more likely (e.g., dancing) and discourages others (e.g.,
conversing), paving the way for the activities themselves to transform the
situation in an emergent process. According to the interactional approach, the
solution to understanding social situations and peoples' experiences lies in an
awareness of the interplay of such situated activities.

Following this interactional perspective, \citet{Uhde2021b} argued that the
``social fit'' of a technology-mediated activity (e.g., a phone call) to
a certain location (e.g., a library) is not so much a consequence of attributes
of the library. Instead, ``library'' should better be understood as shorthand
for a situation in which people read, and noisy ``phone calls'' are usually
incompatible with ``reading''. A more lenient example is a rock concert, which
promotes dancing and shouting. Here, some incompatible activities, such as
reading, are unlikely yet usually not forbidden (although they would seem
a little weird), because they do not strictly conflict with the central concert
activities. But they are uncommon. If a reader finds herself at a concert, she
might bewilder people, and she will have a hard time trying to convince them to
be silent. More generally, \citeauthor{Uhde2021b} argue that such
(in-)compatibilities between co-located activities lead to different types of
situations. In this sense, incompatible activities (e.g., loud chatting in
a library) are not unacceptable per se, but solely \emph{because} of their
incompatibility with other activities (e.g., reading).

In both examples, the library and the rock concert, the location and time imply
an expected activity (i.e., reading and dancing, respectively). In many other
cases, this is less clear and needs to be negotiated ad hoc among the people in
the particular situation. This makes apparent that whether a technology-mediated
activity ``fits'' or not, is neither a matter of the activity, nor of location,
time, people, technology, or environmental attributes as such. Instead, it is
primarily a matter of the emerging obstructing and facilitating relationships
between the activities performed concurrently~\citep{Dourish2004, Uhde2021b}. In
this view, a technology-mediated activity, such as a phone call, is not simply
performed against the backdrop of a surrounding audience and location (i.e.,
``in context''), but in relation to other activities concurrently performed by
many performers. Taken together, the central takeaway from this interactional
perspective is that a closer look at what people do in a certain situation and
how these activities relate to each other gives us a better understanding about
which activities ``fit'', are ``acceptable'', or even appreciated.

\subsection{Technology-Mediated Experiences in Social Situations}

Reeves and colleagues~\citep{Reeves2005} provided an early taxonomy to describe
technology-mediated experiences in social situations. More specifically, they
focused on the experience from the ``spectator's'' perspective (similar to the
``audience'' in Goffman's work or ``observers'' elsewhere,
e.g.,~\citep{Alallah2018, Ens2015}). Reeves and colleagues distinguished between
four main categories of interaction, based on the visibility of their
manipulations and effects to spectators. Interactions with visible manipulations
(e.g., pressing a big red button) and effects (e.g., a light flashes) fall into
the \emph{expressive} category. If only the effects are visible, the interaction
is \emph{magical}. In contrast, if only the manipulation is visible, the
interaction is \emph{suspenseful}. Finally, interactions with invisible
manipulations and effects fall into the \emph{secretive} category. This taxonomy
makes it easier for designers to think about what parts of the interaction
should be revealed or hidden from other people. However, from an interactional
perspective, one of its shortcomings is that it does not account for the
spectator's own activity or other activities performed by surrounding people. As
a consequence, the taxonomy does not capture, for example, whether a spectator
is reading a book or dancing as potentially relevant factors to describe their
experience of someone else's phone call---although we might intuitively expect
very different experiences.  Nonetheless, the model can be useful from an
interactional perspective as well, because it emphasizes that parts of the
interaction need to be at least perceptible to have an effect on others. For
example, in expressive and suspenseful interactions, visible manipulations can
become obtrusive for the spectator's activity and clearly reveal the user as the
``culprit''. In contrast, magical and secretive interactions keep the attention
away from the user.

Empirical comparisons of the four categories found especially suspenseful
interactions (i.e., manipulation visible, effects invisible) to be experienced
as ``awkward'' by observers and recommended using the other forms
instead~\citep{Ens2015, Hakkila2015, Monk2004a, Montero2010}. For example,
\citet{Monk2004a} studied phone calls at bus stops and on trains and found that
observers were annoyed by more suspenseful types of interaction. Note, however,
that the observers' activities were not reported, so we can only tentatively
assume that they were performing relatively independent ``bus stop activities''
(e.g., smoking, waiting) and ``train activities'' (e.g., reading, sleeping).
Montero and colleagues~\citep{Montero2010} asked participants to imagine
different types of interactions (based on the categories by Reeves et
al.~\citep{Reeves2005}) with a smartphone in public places versus at home. They
also found suspenseful interactions to be the least acceptable in public,
although observer activities were again not considered. Finally, Häkkilä and
colleagues~\citep{Hakkila2015} found suspenseful interactions with smart glasses
in a supermarket and on the beach (no observer activities reported) to be
perceived as potentially embarrassing. In sum, suspenseful interactions have led
to the most negative experiences among the four categories in previous studies,
albeit with the restriction that they focused mostly on the user and the
interactive technology, and not on what other people were doing. Consequently,
previous work recommends alternative forms of interaction that keep the
attention away from the user and hide their manipulations (e.g., secretive or
magical interactions~\citep{Koelle2020, Montero2010}).

However, hiding an interaction completely (especially a manipulation) is not
always possible. Thus, a common suggestion to overcome acceptability problems is
to make relatively visible interactions more \emph{subtle}~\citep{Koelle2020,
Rico2010, Pohl2019}. For example, manipulations can take the form of
inconspicuous everyday gestures, such as foot tapping. Subtleness is an attempt
to reduce obtrusiveness (compared to suspenseful and expressive interactions)
while preserving privacy (compared to expressive and magical interactions).
It is the most frequently used strategy in the social acceptability
literature~\citep{Koelle2020}, and it seems to make sense in the locations
studied---supermarkets, streets, and trains. The situations typically related to
these places involve unrelated strangers performing mostly independent
activities. Here, subtle interactions can avoid friction and thus negative
experiences for ``observers''.

But things may be different in other social situations, such as face-to-face
conversations. A conversation is a co-performed activity, which creates
a particular relationship between the conversation partners. Here,
unobtrusiveness and privacy, the central advantages of subtle interactions, may
not be the main concerns of the people involved. After all, they have already
agreed to interact with each other more or less openly. This notion is somewhat
supported by Ahlström and colleagues' \citep{Ahlstrom2014} findings, although
the authors draw different conclusions. They tested spatial hand gestures of
different sizes to interact with a smartphone. The participants felt comfortable
with smaller (i.e., more subtle) hand gestures in front of all ``audiences'',
including strangers. But large (i.e., more suspenseful and potentially more
obtrusive) gestures had high acceptance rates as well, when performed around
family, partners, or friends. We often interact with these people in more
intimate social situations such as face-to-face conversations in enclosed
spaces, and we co-perform many activities. While Ahlström and colleagues
generally recommend the smaller gestures, we would argue that their findings do
not strictly reject larger interactions across all social situations.

Later studies tentatively suggested possible negative side effects of
subtleness. Pohl and colleagues~\citep{Pohl2019} note that subtleness introduces
a risk to be ``uncovered'', which could breach trust in close social
interactions, such as intimate face-to-face conversations. Ens and
colleagues~\citep{Ens2015} presented an alternative design strategy to
communicate the effects of otherwise ``suspicious'' interactions (e.g., with
a smartphone) more openly, for example to communicate that one is not distracted
but looking up relevant information for the current conversation.

Unlike subtle and secretive forms of interaction, expressive and suspenseful
interactions clearly communicate that something is going on. Expressive
interactions additionally communicate their purpose transparently. They are
self-contained and the interaction itself can signal to others that no further
intervention is needed. In contrast, suspenseful interactions can be more
difficult for observers to interpret. They may require more prior knowledge
about the interaction. This can be problematic in situations where people
perform independent activities among strangers, which has possibly contributed
to the general advice against them~\citep{Ens2015, Koelle2020}. But in closer,
more direct social interactions, the observers (who are more involved) might
already have more knowledge about the user and the technology including its
effects, or at least they have no reason to suspect anything harmful.

In sum, the standard recommendation for subtle and unobtrusive forms of
interaction, and against suspenseful interactions, seems at least in part to
result from the mostly anonymous, public situations chosen in previous studies.
We believe that this may have contributed to the current focus on the
\emph{form} of ``socially acceptable'' interactions, instead of a focus on their
situational meaning, which can vary depending on the social situation at hand.
Pohl and colleagues~\citep{Pohl2019} assumed that subtle interactions may have
negative effects in face-to-face conversations. Additionally, the negative
consequences of suspenseful interactions may be reduced if the observer has some
prior knowledge about the interaction, and if they are performed as part of
a safe social co-performance, such as a conversation. To show this, we carried
out an experimental study in which we compared the experiential consequences of
subtle and suspenseful forms of interaction in such a face-to-face conversation
situation.

\section{Study: Subtle and Suspenseful Interactions in Face-to-face Conversations}

For our experimental study, we selected face-to-face conversations as the type
of situation in which we assumed a negative effect of subtle interactions (based
on~\citep{Pohl2019}), compared to the usually not recommended suspenseful
interactions (e.g.,~\citep{Montero2010, Koelle2020}). We focused on the observer
experience of an interaction with a hearing aid prototype during these
conversations. This setting and device seemed appropriate for several reasons.
First, as outlined above, face-to-face conversations imply a direct, more
trustful social interaction between performer and observer, which is different
from the relatively anonymous co-located settings on a train or in a supermarket
studied before. In addition, they are relatively easy to set up in a controlled
setting, compared to, for example, a romantic date. Second, we chose a hearing
aid as our interactive technology, because it represents an essential element of
the conversation. The hearing aid enables the user to listen and participate,
and it does not introduce unrelated distractions. It also naturally hides the
``effects'' of the interaction (i.e., improved hearing), which is consistent
both with subtle and suspenseful forms of interaction.

We compared the experiential effects of subtle vs.\ suspenseful interactions
vs.\ a control condition [no interaction] in an experimental between-group
design. To date, there is no standard measure for ``social
acceptability''~\citep{Koelle2020}, and instead we attempted to capture the
situated experience from the observer perspective as comprehensively as
possible. We measured several experiential aspects of the conversation,
including the participants' (observers) subjective emotional experience, their
perceived quality of the conversation, their impression of the conversation
partner (our confederate), and how they perceived and evaluated the interactive
technology.

Our hypotheses were:

\begin{description}
  \item [H1:] Subtle interactions lead to a more negative affective experience
    of the conversation, compared to suspenseful interactions.
  \item [H2:] Subtle interactions lead to a more negative perception of the
    quality of the conversation, compared to suspenseful interactions.
  \item [H3:] Subtle interactions lead to a more negative product perception of
    the interactive technology, compared to suspenseful interactions.
\end{description}

\noindent{}In addition, we also tested whether the form of interaction had an
effect on the perception of the conversation partner's personality, but without
directed hypotheses.

\subsection{Participants}

We recruited 36 people through an agency, one of whom did not appear to the
study. Thus, we had 35 participants (18 female, median age = 40, ranging from 30
to 52) with a wide variety of backgrounds (e.g., scaffolder, pharmaceutical
researcher, engineer, event manager, teacher, pediatric nurse, etc.). In
addition, we asked the agency for participants without a hearing impairment to
avoid strong differences between participants' prior experiences with hearing
aids. They were compensated with 30€.

\subsection{Procedure}

The experiment took place on the premises of a market research facility located
in Germany. Each session lasted about 30 minutes and consisted of four parts:
(1) introduction by the examiner, (2) the conversation with the confederate, (3)
questionnaires, and (4) a debriefing.

\subsubsection{Introduction}

The examiner welcomed each participant separately in the lobby of the market
research facility. He announced that the first step of the study is to have an
approximately seven-minute conversation with another researcher (i.e., the
confederate) about hobbies. This other researcher would already be waiting in
a separate room. The examiner also told the participant that the conversation
partner had a hearing loss and that he would be wearing a hearing aid, which,
however, provides him with a full compensation of his hearing. The examiner
further announced that the conversation would be recorded. Before the
conversation, the examiner asked the participant to wait for another moment and
excused himself. He went into the interview room, supposedly to check whether
everything is ready. Inside the room he activated two time-controlled noise
distractions as well as the recording equipment. Upon re-entering the lobby, he
informed the participant that everything is ready.

\subsubsection{Conversation}

The participant went into the conversation room, where the researcher (i.e., the
confederate actor) welcomed them and asked them to take a seat opposite of him.
The researcher started the conversation with a short introduction to his
hobbies. After finishing, he asked ``and you?'' to start the conversation. To
keep the conversation going, the researcher had a repertoire of eight questions
addressing the ``what, how, and why'' of the participant's hobbies. While the
conversation took place, we played back some soft background noises resembling
muted sound from a larger group of people in a neighboring room. At minute 02:56
and 04:53, we staged two time-controlled audio distractions. At minute 02:56,
a man's voice became noticeable, he talked a bit louder, sounding slightly
furious for about 20 seconds. This distraction came from the same direction as
the background noise. At 04:53, a mobile phone rang for about 20 seconds out of
a different direction than the background noise. The researcher always briefly
acknowledged the noise, but did not stop the conversation. Depending on the
experimental condition, the researcher performed either a suspenseful
interaction (open hand gesture), a subtle interaction (using a small handheld
device) or no interaction (control condition) to supposedly refocus his hearing
aid from the noise back to the conversation partner. Note that the researcher
neither mentioned nor explained this interaction to the participant. In fact,
none of the participants commented on this interaction, and often they did not
even pause the conversation. After about seven minutes, the researcher thanked
the participant for the stimulating conversation and took her or him to another
room to fill in a number of questionnaires (details described below).

\subsubsection{Debriefing}

After completing the questionnaires, we carried out a short debriefing session.
During this session we disclosed the true purpose of the study and the fact that
the conversation partner did not actually have a hearing loss. We answered all
related questions the participants had about the study.

\begin{table*}[t]
	\centering
	\caption{Spearman correlations of all measures included in our study. Asterisks mark significant correlations (* = 5\% level; ** = 1\% level). The values in brackets are the internal consistencies (Cronbach's $\boldsymbol{\alpha}$) of the respective scales. PA = Positive Affect, NA = Negative Affect, O = Openness, C = Conscientiousness, E = Extraversion, A = Agreeableness, N = Neuroticism, HQ = Hedonic Quality, PQ = Pragmatic Quality}
    \footnotesize%
\label{tab:correlations}
	{
		\begin{tabular}{ld{1.4}d{1.4}d{1.4}d{1.4}d{1.4}d{1.4}d{1.4}d{1.4}d{1.4}d{1.4}d{1.4}d{1.4}}
			\toprule
			Measure           & \multicolumn{1}{c}{Conversation}
			                  & \multicolumn{1}{c}{PA}
			                  & \multicolumn{1}{c}{NA}
			                  & \multicolumn{1}{c}{O}
			                  & \multicolumn{1}{c}{C}
			                  & \multicolumn{1}{c}{E}
			                  & \multicolumn{1}{c}{A}
			                  & \multicolumn{1}{c}{N}
			                  & \multicolumn{1}{c}{HQ}
			                  & \multicolumn{1}{c}{PQ}
			                  & \multicolumn{1}{c}{Beauty} \\
			\midrule
			Conversation      & (.89)        &         &         &          &           &           &         &       &          &          &         \\
			Positive Affect   & 0.68$**$     & (.77)   &         &          &           &           &         &       &          &          &         \\
			Negative Affect   & -0.10        & -0.05   & (.71)   &          &           &           &         &       &          &          &         \\
			Openness          & 0.20         & 0.28    & 0.03    & (.29)    &           &           &         &       &          &          &         \\
			Conscientiousness & 0.30         & 0.37$*$ & 0.05    & 0.10     & (.61)     &           &         &       &          &          &         \\
			Extraversion      & 0.43$*$      & 0.39$*$ & -0.01   & 0.43$*$  & 0.52$**$  & (.72)     &         &       &          &          &         \\
			Agreeableness     & 0.29         & 0.10    & -0.21   & 0.40$*$  & 0.11      & 0.35$*$   & (.31)   &       &          &          &         \\
			Neuroticism       & -0.19        & -0.33   & -0.18   & -0.30    & -0.51$**$ & -0.58$**$ & -0.06   & (.08) &          &          &         \\
			HQ                & 0.43$*$      & 0.37$*$ & 0.12    & 0.42$*$  & -0.07     & 0.53$**$  & 0.31    & -0.18 & (.84)    &          &         \\
			PQ                & 0.39$*$      & 0.30    & 0.12    & 0.38$*$  & -0.03     & 0.29      & 0.34$*$ & -0.18 & 0.77$**$ & (.80)    &         \\
			Beauty            & 0.36$*$      & 0.19    & 0.27    & 0.22     & -0.03     & 0.34$*$   & 0.18    & -0.22 & 0.64$**$ & 0.55$**$ &         \\
			Goodness          & 0.34         & 0.19    & 0.04    & 0.47$**$ & 0.03      & 0.50$**$  & 0.33    & -0.12 & 0.77$**$ & 0.82$**$ & 0.43$*$ \\
			\bottomrule
		\end{tabular}
	}
\end{table*}

\subsection{Material and Methods}

\subsubsection{Hearing Aid}

\begin{figure}[b]
  \centering
  \includegraphics[width=0.49\linewidth]{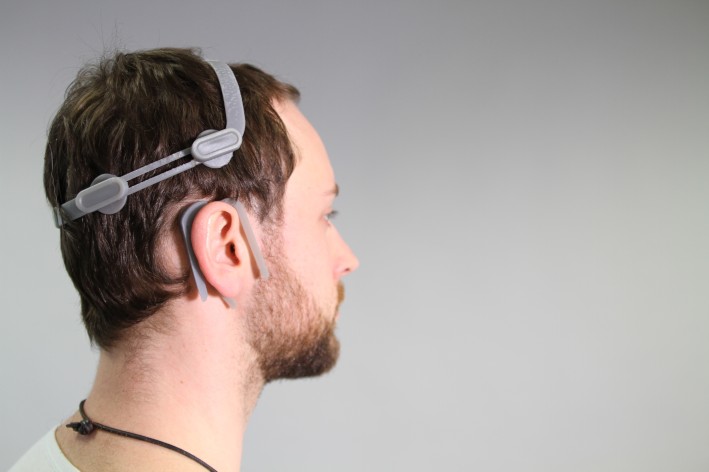}
  \includegraphics[width=0.49\linewidth]{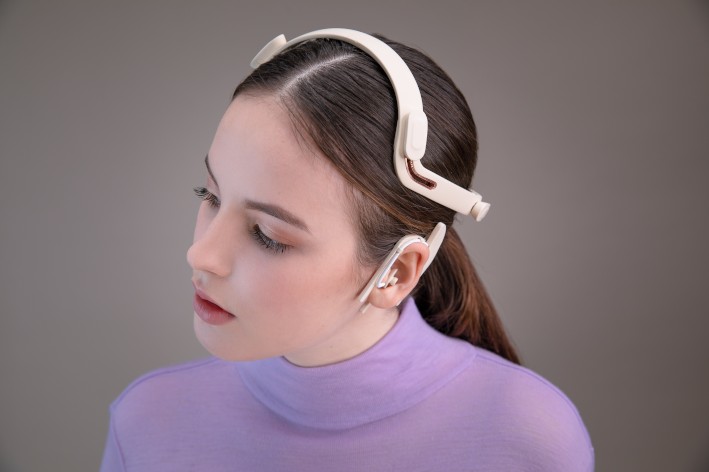}
  \caption{The prototypical hearing aid used in our study (left) and the
           envisioned final design (right).}
  \Description{Left: The back of the head of a man wearing a hearing aid that
  spans across the skull and around the ear. Right: A young woman wearing a more
  polished design of the same hearing aid.}%
\label{fig:hearing_aid}
\end{figure}

Figure~\ref{fig:hearing_aid} shows the hearing aid used in the study. It is
a non-functional design prototype from a past project that is supposed to be
controlled by a brain-computer interface as well as gesture-based interaction.
Note that, although we used this hearing aid as a vehicle to study the impact of
interaction forms on social perceptions in a realistic scenario, the particular
aid and its technical features were neither explained to nor discussed with the
participants at any time other than the debriefing.

The envisioned hearing aid has a ``beam forming'' function, which means that it
can adjust the focus of hearing in a certain direction, both automatically and
manually. For example, it can amplify the voice of a conversation partner and
fade out environmental noise. However, this function sometimes erroneously
focuses on the wrong sound source and accidentally fades out the conversation
partner's voice instead. In such cases, a manual intervention is needed, for
which the subtle and suspenseful interactions are used. Our confederate actor
used these gestures as reactions to the two staged incidents during the
conversation, where the hearing aid supposedly focused on the wrong sound
source.

\subsubsection{Two Forms of Interaction}

The interactions used in the study were drawn from a set of 28 gestures
developed in an earlier workshop. We invited an actor to design suitable
free-form gestures for controlling different functions of the hearing aid. The
actor was led through eight scenarios, ranging from ``having brief discussions
with a group of people'' to ``walking down a busy street while talking and
monitoring cars.'' For each scenario, the actor was provided with a reason to
manipulate the hearing aid, such as ``change focus from person 1 to person
2 manually.'' The actor was then asked to imagine the scenario and to act out
aesthetic, meaningful, and discernible gestures to invoke the particular
function (in the sense of Bodystorming, e.g.,~\citep{Oulasvirta2003}).

\begin{figure}[b]
  \centering
  \includegraphics[width=\linewidth]{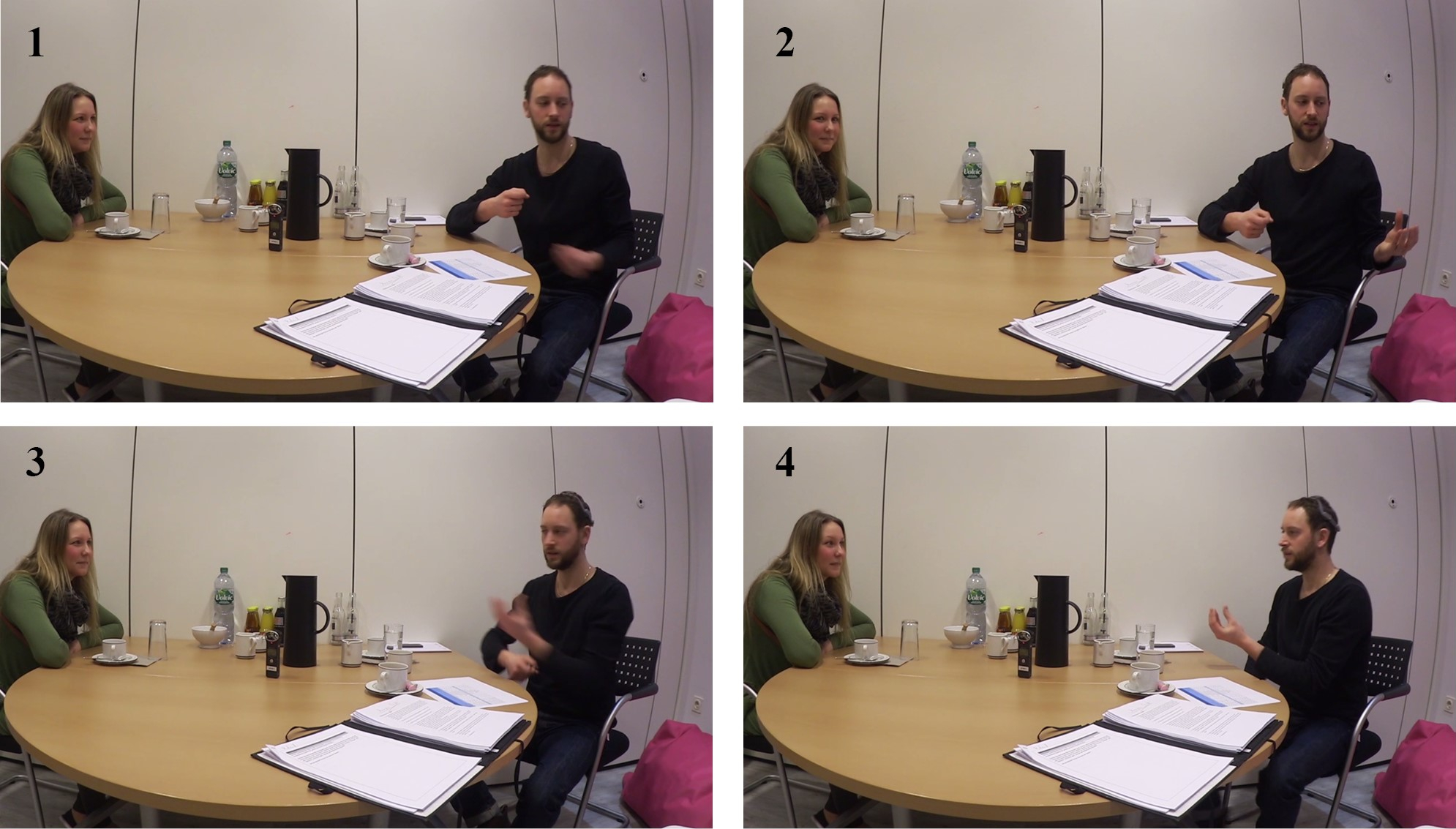}
  \caption{The suspenseful interaction from our study.}
  \Description{Four photos from the study illustrating the suspenseful
  interaction as described in the text. They depict the researcher (our
  confederate on the right side of a table, and the participant on the left,
  opposite of the researcher.)}%
\label{fig:gesture_interaction}
\end{figure}

\begin{figure}[b]
  \centering
  \includegraphics[width=\linewidth]{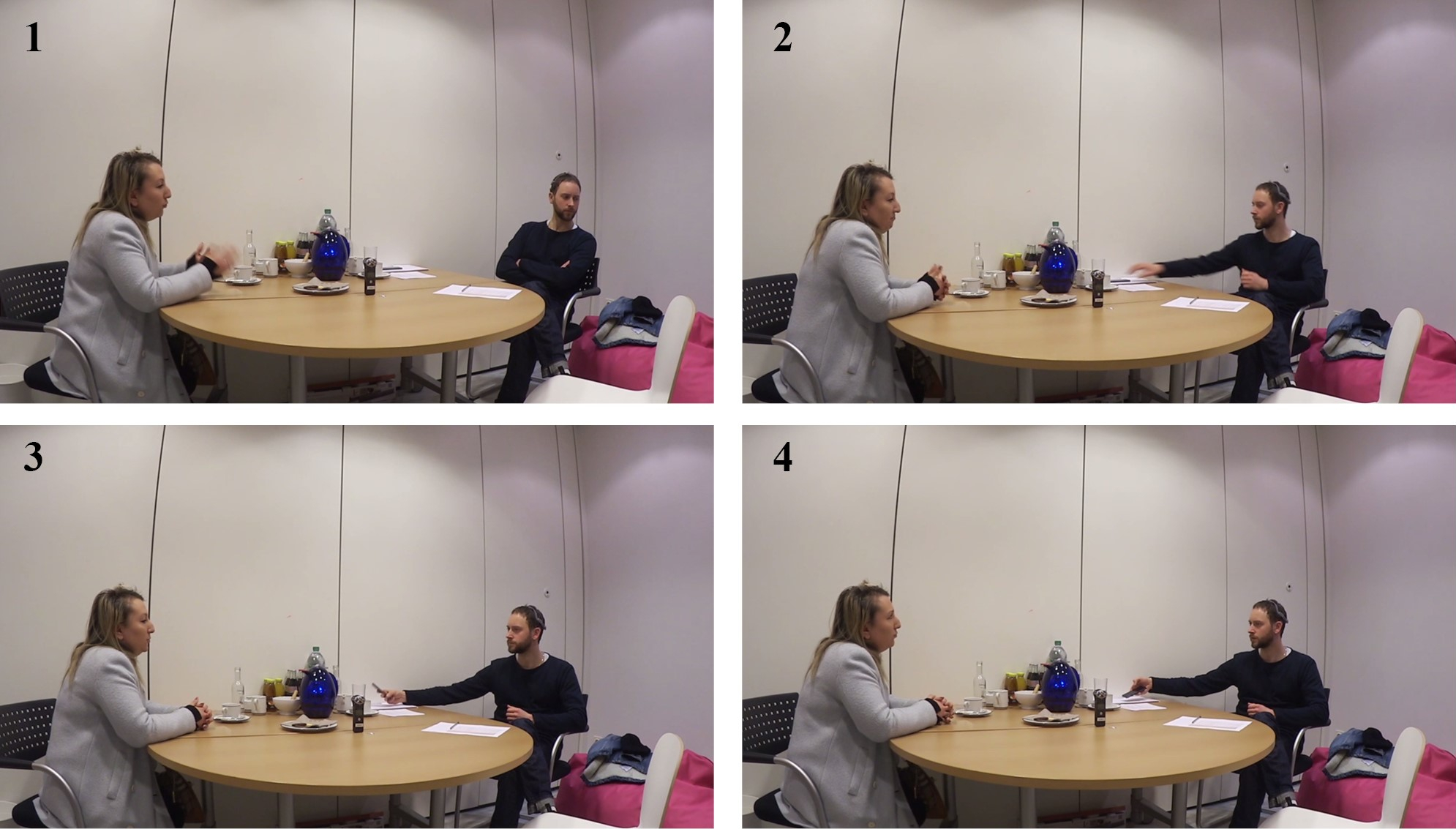}
  \caption{The subtle interaction from our study.}
  \Description{Four photos from the study illustrating the subtle interaction as
  described in the text.}%
\label{fig:subtle_interaction}
\end{figure}

From this pool, we selected two gestural interactions for our face-to-face
conversation scenario. The first interaction was a spacious, ``suspenseful''
hand gesture that was highly visible to the observer. It started with orienting
the upper body towards the unwanted sound source. The researcher then moved the
closed hand into the direction of the noise. The hand was lifted and opened, and
then directed towards the participant. He then lowered the hand again and closed
it (see Figure~\ref{fig:gesture_interaction}). The second interaction was
a ``subtle'' gesture, smaller and less obtrusive, based on an interaction with
a handheld device (as recommended for example in~\citep{Koelle2020, Rico2010}).
The researcher picked up the device and used his thumb to quickly draw a ``V''
shape on its surface. He then tilted the device in the direction of the
participant and placed it back on the table (see
Figure~\ref{fig:subtle_interaction}). This lasted for about three seconds. The
whole gesture was performed while keeping eye contact with the conversation
partner. The effect of the interaction in terms of Reeves and colleagues'
taxonomy~\citep{Reeves2005} was hidden in both cases.
h
\subsubsection{Questionnaires}

After the conversation, we asked the participants about (1) their affective
experience, (2) their perceived quality of the conversation, (3) their
impression of the conversation partner, and (4) their perception and evaluation
of the interactive hearing aid. Correlations between measures and internal
consistencies are reported in Table~\ref{tab:correlations}.

To measure affective experiences, we used the short, ten item version of the
Positive and Negative Affect Schedule (PANAS)
questionnaire~\citep{Mackinnon1999, Watson1988} in its German translation
\citep{Krohne1996}. We computed positive affect (PA) by averaging the responses
to the items: ``alert'', ``determined'', ``enthusiastic'', ``excited'', and
``inspired'' per person, ranging from ``not at all'' (1) to ``extremely'' (5).
We computed negative affect (NA) by averaging the responses to the items:
``afraid'', ``upset'', ``scared'', ``nervous'', and ``distressed'' per person.
Internal consistencies were good for both scales (see
Table~\ref{tab:correlations}).

To assess the perceived quality of the conversation, participants rated how
``interesting'', ``informative'', ``pleasant'', and ``positive'' the
conversation had been on 7-point items ranging from ``not at all'' (1) to
``extremely'' (7). The internal consistency was high.

Subsequently, we asked participants to assess the conversation partner's
personality in terms of the Big Five personality traits (i.e., neuroticism,
extraversion, openness to experience, agreeableness, conscientiousness). Typical
personality questionnaires are relatively long (e.g., the
NEO-PI-R~\citep{Berth2006} with 240 items or the Big Five Inventory (BFI-44)
with 44 items), and instead we opted for the short version of the Big Five
Inventory with only ten items (BFI-10; \citep{Rammstedt2007}). The BFI-10
includes two items for each personality trait. We slightly rephrased the wording
to fit with an other-assessment from the observer perspective. For example,
extraversion was measured with the two items: ``I see my conversation partner as
someone who is outgoing, sociable'' and ``I see my conversation partner as
someone who is reserved'' (inverted). Each item was measured with a five point
scale ranging from ``disagree strongly'' (1) to ``agree strongly'' (5), which we
then averaged per person. However, partially due to the few items and relatively
small sample size, internal consistencies were low for three of the personality
dimensions, borderline acceptable for ``conscientiousness'', and satisfactory
only for ``extraversion'' (see Table~\ref{tab:correlations}).

To assess the observers' perception of the interactive device, we included the
AttrakDiff Mini questionnaire \citep{Hassenzahl2010c}. It consists of two scales
to capture hedonic (HQ) as well as pragmatic quality perceptions (PQ). Note that
in our case the hearing aid only differed in terms of the form of interaction
between conditions. The HQ scale consists of the four 7-point semantic
differential items ``stylish-tacky'' (inverted), ``cheap-premium'',
``unimaginative-creative'' and ``dull-captivating''. The PQ scale consists of
the four 7-point semantic differential items ``simple-complicated'' (inverted),
``practical-impractical'' (inverted), ``predictable-unpredictable'' (inverted)
and ``confusing-structured''. The general evaluation (``goodness'') is measured
with the single item ``good-bad'' (inverted), and perceived beauty with the item
``ugly-beautiful''.

\section{Results}

\begin{figure*}[t]
\centering
\includegraphics[width=\linewidth]{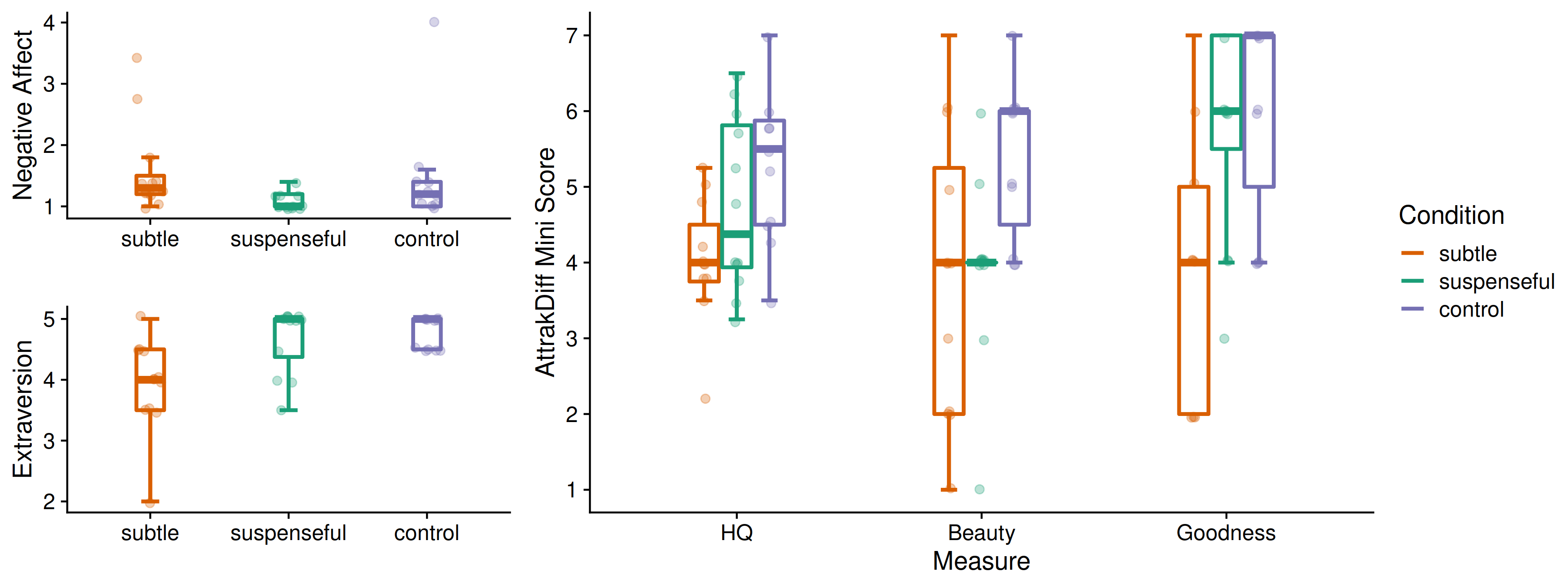}
\caption{Summary plot for all measures with statistically significant
  differences between the three conditions. The scales for Negative Affect and
  Extraversion are slightly shortened to improve visibility while still
  including all data points. HQ = Hedonic Quality}
\Description{Three separate box plots of the differences between the three
  experimental conditions in terms of 1) Negative Affect, 2) Extraversion,
  and 3) Product Perception. The first box plot shows the slightly stronger
  negative affect in the subtle condition, on a scale from 1 to 4. Except for two
  outliers in the subtle condition and one in the control condition, all values
  remain between 1 and 2. The second box plot shows the lower rating of the
  conversation partner's extraversion on a scale from 1 to 5 in the subtle
  condition. The subtle condition has the largest spread from around 2 to
  around 5, followed by the suspenseful condition from around 3.5 to 5, and all
  values in the control condition are between around 4.5 and 5. The third
  box plot shows the lower score of the subtle condition on the AttrakDiff Mini
  scales (hedonic quality, beauty, goodness) on a scale from 1 to 7. For hedonic
  quality, the subtle condition has an outlier slightly above 2, and most values
  range between 3.5 and 5.5. The suspenseful and control conditions have a larger
  spread from around 3.5 to 6.5, and from around 3.5 to 7, respectively. For
  beauty, the subtle condition spreads across the whole scale from 1 to 7,
  suspenseful has a minimal box at 4 with two outliers to each side, and all
  control condition scores range from around 4 to 7. Finally, for goodness, the
  subtle condition has the largest spread from around 2 to 7. Suspenseful and
  control conditions both spread from around 4 to 7, with one outlier at around 3
  in the suspenseful condition.}%
\label{fig:boxplots}
\end{figure*}

Before our analysis, we ran assumption checks and found several non-normal
distributions across our dependent measures. This was in part due to the small
cell sizes that were a consequence of our expensive test setup involving
a professional actor (typically 30 participants or more per cell are
recommended, see e.g.,~\citep[][p. 700]{Field2017}). During a visual exploratory
analysis, we also found some outliers (e.g., for negative affect, hedonic
quality, and goodness, see Figure~\ref{fig:boxplots}). Thus, we decided to run
all of our analyses using non-parametric methods for hypothesis testing that
make no normality assumption and are robust against outliers. This includes the
correlations reported in Table~\ref{tab:correlations} and summary statistics
reported in Table~\ref{tab:summary}. Box plots for all hypothesis tests with
significant results can be found in Figure~\ref{fig:boxplots}.

\subsection{Affective Experience of the Conversation}

A Kruskal-Wallis test with the type of interaction (suspenseful, subtle,
control) as factor and negative affect (NA) as measure revealed a significant
effect ($H = 6.49$; $\text{df} = 2$; $p < .05$). Pairwise
comparisons\footnote{For pairwise comparisons here and in later tests we used
the method proposed by Siegel and Castellan~\citep{Siegel1988} using the
R package pgirmess~\citep{Giraudoux2013}. The p values for post hoc pairwise
comparisons not based on our hypotheses (e.g, everything involving the control
condition) are adjusted for family-wise error.} confirmed that this was caused
by higher negative affect in the subtle condition, compared to the suspenseful
condition ($\text{diff}_{observed} = 10.05$; $\text{diff}_{critical} = 8.15$; $p
< .05$). On the item level, this difference was based on increased ``distress''
and ``nervousness'' in the subtle interaction condition. The differences between
the suspenseful and the control condition, and between subtle and control were
not significant. A similar test with positive affect (PA) was not significant
($H = 2.26$; $\text{df} = 2$; $p = .32$).

Thus, the subtle interaction led to higher negative affect, compared to the
suspenseful interaction. This confirms our hypothesis H1 on the overall
affective experience for negative affect. An effect on positive affect could not
be found.

\subsection{Perception of the Conversation}

Next, we ran a Kruskal-Wallis test to compare the effect of the type of
interaction (suspenseful, subtle, control) on the perceived quality of the
conversation. However, there was no significant effect ($H = 2.53$; $\text{df}
= 2$; $p = .28$). Thus, hypothesis H2 could not be confirmed: The type of
interaction did not influence the perceived quality of the conversation.

\subsection{Product Perception and Evaluation}

A Kruskal-Wallis test with type of interaction (suspenseful, subtle, control) as
factor and hedonic quality perception (HQ) as measure revealed a significant
effect ($H = 6.71$; $\text{df} = 2$; $p < .05$). This was caused by a lower
perceived hedonic quality in the subtle interaction, compared to the control
condition ($\text{diff}_{observed} = 10.95$; $\text{diff}_{critical} = 10.17$;
$p_{adjusted} < .05$). There was no difference between the suspenseful
interaction and the control condition or between the suspenseful and subtle
interaction. A similar Kruskal-Wallis test with perceived pragmatic quality as
measure revealed no significant effect ($H = 3.88$; $\text{df} = 2$; $p = .14$).

Next, we conducted a Kruskal-Wallis test with perceived product beauty as
measure and found a significant effect ($H = 7.61$ $\text{df} = 2$; $p < .05$).
However, no pairwise comparison revealed a significant difference between
specific groups after adjustments for family-wise error. Finally, we ran
a Kruskal-Wallis test with overall product perception (``goodness'') as measure
and found a significant effect ($H = 7.04$; $\text{df} = 2$; $p < .05$). The
overall product perception was better in the suspenseful condition than in the
subtle condition ($\text{diff}_{observed} = 8.15$; $\text{diff}_{critical}
= 8.11$; $p < .05$), and better in the control condition than in the subtle
condition ($\text{diff}_{observed} = 10.34$; $\text{diff}_{critical} = 10.09$;
$p_{adjusted} < .05$). The suspenseful and control conditions were not
significantly different.

Taken together, the subtle interaction led to a significantly more negative
product perception than the suspenseful interaction and the control condition.
The effect was strongest for the overall goodness, and partially confirms H3.
Hedonic quality was also significantly more negative in the subtle interaction
condition, compared to the control condition.

\begin{table}[t]
	\centering
	\caption{Summary statistics for all measures in each condition. MAD = Median Adjusted Difference; IQR = Interquartile Range}
    \tiny%
\label{tab:summary}
	{
		\begin{tabular}{lld{1}d{1}d{1}d{1}d{1}d{1}}
			\toprule
			                  &
			                  & \multicolumn{1}{c}{Median}
			                  & \multicolumn{1}{c}{MAD}
			                  & \multicolumn{1}{c}{IQR}
			                  & \multicolumn{1}{c}{Range}
			                  & \multicolumn{1}{c}{Minimum}
			                  & \multicolumn{1}{c}{Maximum} \\
			\cmidrule[0.4pt]{1-8}
			Positive Affect   & control     & $3.60$ & $0.60$ & $1.10$ & $3.20$ & $1.60$ & $4.80$  \\
			                  & subtle      & $3.40$ & $0.40$ & $0.50$ & $2.60$ & $1.40$ & $4.00$  \\
			                  & suspenseful & $3.50$ & $0.50$ & $0.95$ & $3.25$ & $1.75$ & $5.00$  \\\addlinespace
			Negative Affect   & control     & $1.20$ & $0.20$ & $0.40$ & $3.00$ & $1.00$ & $4.00$  \\
			                  & subtle      & $1.30$ & $0.10$ & $0.30$ & $2.40$ & $1.00$ & $3.40$  \\
			                  & suspenseful & $1.00$ & $0.00$ & $0.20$ & $0.40$ & $1.00$ & $1.40$  \\\addlinespace
			Conversation      & control     & $7.00$ & $0.00$ & $1.38$ & $3.75$ & $3.25$ & $7.00$  \\
			                  & subtle      & $5.75$ & $1.00$ & $1.75$ & $5.00$ & $2.00$ & $7.00$  \\
			                  & suspenseful & $6.46$ & $0.54$ & $1.13$ & $1.50$ & $5.50$ & $7.00$  \\\addlinespace
			Agreeableness     & control     & $4.00$ & $0.50$ & $0.75$ & $3.00$ & $2.00$ & $5.00$  \\
		                      & subtle      & $4.00$ & $0.25$ & $0.50$ & $1.50$ & $3.00$ & $4.50$  \\
			                  & suspenseful & $4.50$ & $0.50$ & $1.63$ & $3.00$ & $2.00$ & $5.00$  \\\addlinespace
			Extraversion      & control     & $5.00$ & $0.00$ & $0.50$ & $0.50$ & $4.50$ & $5.00$  \\
			                  & subtle      & $4.00$ & $0.50$ & $1.00$ & $3.00$ & $2.00$ & $5.00$  \\
			                  & suspenseful & $5.00$ & $0.00$ & $0.63$ & $1.50$ & $3.50$ & $5.00$  \\\addlinespace
			Conscientiousness & control     & $4.50$ & $0.50$ & $1.00$ & $2.00$ & $3.00$ & $5.00$  \\
			                  & subtle      & $4.00$ & $0.50$ & $0.63$ & $1.50$ & $3.50$ & $5.00$  \\
			                  & suspenseful & $4.50$ & $0.50$ & $0.50$ & $2.50$ & $2.50$ & $5.00$  \\\addlinespace
			Neuroticism       & control     & $1.50$ & $0.50$ & $1.00$ & $2.50$ & $1.00$ & $3.50$  \\
			                  & subtle      & $2.50$ & $0.25$ & $0.63$ & $2.00$ & $1.00$ & $3.00$  \\
			                  & suspenseful & $1.75$ & $0.50$ & $0.75$ & $2.00$ & $1.00$ & $3.00$  \\\addlinespace
			Openness          & control     & $4.00$ & $0.50$ & $1.00$ & $2.50$ & $2.50$ & $5.00$  \\
		          	          & subtle      & $3.50$ & $0.50$ & $0.63$ & $2.50$ & $2.50$ & $5.00$  \\
		        	          & suspenseful & $4.00$ & $0.50$ & $0.75$ & $2.00$ & $3.00$ & $5.00$  \\\addlinespace
			Hedonic Quality   & control     & $5.50$ & $1.00$ & $1.38$ & $3.50$ & $3.50$ & $7.00$  \\
		                      & subtle      & $4.00$ & $0.25$ & $0.75$ & $3.00$ & $2.25$ & $5.25$  \\
		                      & suspenseful & $4.38$ & $0.88$ & $1.88$ & $3.25$ & $3.25$ & $6.50$  \\\addlinespace
			Pragmatic Quality & control     & $6.25$ & $0.75$ & $1.38$ & $3.25$ & $3.75$ & $7.00$  \\
		                      & subtle      & $4.75$ & $1.38$ & $2.50$ & $4.50$ & $2.50$ & $7.00$  \\
			                  & suspenseful & $5.25$ & $0.63$ & $1.50$ & $4.50$ & $2.25$ & $6.75$  \\\addlinespace
			Beauty            & control     & $6.00$ & $1.00$ & $1.50$ & $3.00$ & $4.00$ & $7.00$  \\
			                  & subtle      & $4.00$ & $2.00$ & $3.25$ & $6.00$ & $1.00$ & $7.00$  \\
			                  & suspenseful & $4.00$ & $0.00$ & $0.00$ & $5.00$ & $1.00$ & $6.00$  \\\addlinespace
			Goodness          & control     & $7.00$ & $0.00$ & $2.00$ & $3.00$ & $4.00$ & $7.00$  \\
			                  & subtle      & $4.00$ & $2.00$ & $3.00$ & $5.00$ & $2.00$ & $7.00$  \\
			                  & suspenseful & $6.00$ & $1.00$ & $1.50$ & $4.00$ & $3.00$ & $7.00$  \\
			\bottomrule
		\end{tabular}
	}
\end{table}

\subsection{Perception of the Conversation Partner}

Finally, we ran a Kruskal-Wallis test with type of interaction (suspenseful,
subtle, control) as factor and extraversion as measure and found a significant
effect ($H = 12.05$; $\text{df} = 2$; $p < .01$). The conversation partner
appeared less extraverted in the subtle condition than in the suspenseful
condition ($\text{diff}_{observed} = 11.58$; $\text{diff}_{critical} = 10.01$;
$p_{adjusted} < .05$). Additionally, he appeared less extraverted in the subtle
condition than in the control condition ($\text{diff}_{observed} = 12.42$;
$\text{diff}_{critical} = 10.24$; $p_{adjusted} < .05$). There was no difference
between the suspenseful and control conditions. Similar analyses with the
remaining four personality traits revealed no significant effects ($p_{openness}
= .07$; $p_{conscientiousness} = .09$; the other two $p > .1$).

In sum, the subtle interaction made the hearing aid user appear less
extraverted, compared to the suspenseful interaction and control condition.

\section{Discussion}


In the present study, we compared the impact of subtle versus suspenseful forms
of interaction on the observer experience during face-to-face conversations.
Specifically, we used interactions with a hearing aid as our test case and found
negative effects of a subtle form of interaction, compared to a ``suspenseful''
form of interaction and the control condition (no interaction). The subtle form
of interaction had negative effects on the emotional experience of the observer
and their perception and evaluation of the technology itself. In addition,
subtle interactions also led to impressions of the conversation partner as less
extraverted, compared to the other two conditions. Taken together, the type of
interaction had a profound impact on the observer experience, and the subtle
interaction had a consistently more negative effect than the suspenseful
interaction. These findings contradict current social acceptability design
recommendations to use unobtrusive, subtle design strategies and to avoid
``suspenseful'' forms of interaction~\citep{Koelle2020}, and represent the main
contribution of this paper.

As the second contribution, we take this empirical inconsistency as a prompt to
reflect on the current, relatively standardized recommendations for subtle and
against suspenseful forms of interaction. Please note that we do not suggest
that previous studies where subtle interactions performed well are in some form
``invalid'' or that we consider subtleness generally as a ``bad'' design
strategy. Instead, we want to encourage researchers and designers to emphasize
situational differences and to focus their attention more towards situated
meaning, that is, on \emph{why} a certain form of interaction may or may not be
appropriate in a given situation, before designing the form itself (i.e., the
\emph{how}).

\subsection{Limitations}

Before we discuss the broader impact of our study, we would like to highlight
its major limitations. First, some of our scales to measure personality traits
had relatively low internal consistencies, which makes their interpretation
difficult. Statistical reasons for this include that the scales used to measure
each personality trait were quite short (2 items each), and that our sample was
relatively small. Another possible reason may relate to the specific personality
traits themselves and our study situation. Some traits, such as neuroticism and
conscientiousness, may be especially difficult to deduce, based on a short
conversation with a stranger. For future work, we would suggest using more
focused but longer scales to measure certain personality traits of interest.

Second, the chosen device, an interactive hearing aid, and the specific setup,
a face-to-face conversation, may have led to particular, situation-specific
effects that cannot easily be generalized to other situations, such as
interacting with a smartphone on the train. However, this limitation actually
supports the more general point we make. A key takeaway from this work should be
that we cannot simply transfer findings from an interaction in one situation to
another interaction in another situation. Further below, we try to clarify under
which circumstances we believe that findings can be transferred to an extent,
and how we would describe similarities and differences between situations that
are essential for the transferability.

Third, the sample size (35) was relatively small, which was in part due to the
expensive setup involving a professional actor. We chose this setup because we
wanted to create an immersive social situation with realistic experiences of
a face-to-face conversation. Future studies using for example a larger sample
size in an online study could further extend the findings.

\subsection{Reflections on Social Acceptability and Situatedness}

Given the tension between our findings and previous work, we use the remainder
of this paper to reflect on interacting in social situations and social
acceptability. This reflection follows the interactional approach to social
situatedness outlined above (and e.g., in~\citep{Dourish2004, Uhde2021b}).

\subsubsection{Acknowledging Formative Effects and Shared Experiences}

In previous social acceptability studies, the other people around a user have
typically not been the focus of attention. Conceptually, they have been framed
as rather passive ``elements'' of the social situation through labels such as
``audience''~\citep{Rico2010}, ``observer''~\citep{Alallah2018}, and
``spectator''~\citep{Reeves2005}. All of these focus on their role as people who
look at a performed interaction in some form, but not as active contributors on
their own. However, people have an immense formative power to shape social
situations through their activities, such as people dancing in a club who make
it easier for others to dance along. So far, we do not know how the social
acceptability of an interaction relates to the activities performed by other
people.

Perhaps even more strikingly, the interactions themselves can be read as
positioned with a somewhat ``passive'' notion. If we label an interaction as
socially ``acceptable'', we imply the judgement of some authority that can
accept or reject it. The active role of an interaction to shape a situation and
maybe redefine what is considered acceptable in that situation (e.g., the first
dancers on the dance floor) should be studied further.

\subsubsection{Suspenseful Interactions in the Wild}

In our study, the form of interaction in the ``suspenseful'' condition was quite
large and gesture-based, which was previously deemed socially unacceptable
(e.g.,~\citep{Ahlstrom2014, Montero2010}). But despite these recommendations,
suspenseful interactions are actually quite common in real-world social
situations.  For example, playing a smartphone game such as ``Fruit Ninja'' or
taking a photo are usually ``suspenseful'' interactions. Some of them are
directed at a device, which may increase acceptability~\citep{Koelle2020}. But
in terms of the interaction taxonomy, only the manipulations, but not the
effects, are visible---which makes these interactions suspenseful. In fact, some
design interventions such as the ``manner mode'' (silent mode) for smartphones
have been developed and are sometimes enforced~\citep{Srivastava2005} to reduce
the loudness or visibility of ``effects'' (e.g., ringtones, touch-tones), thus
deliberately transforming expressive into suspenseful interactions.

This seems inconsistent with previous recommendations against suspenseful
interactions (e.g.,~\citep{Montero2010, Monk2004a, Ens2015}). But from our
perspective, their wide adoption in practice indicates that they may in fact not
be as generally ``unacceptable'' as believed, and our study adds experimental
findings that support this view. In sum, we find that suspenseful interactions
do not seem unacceptable in general, but the experiences they create depend
more specifically on the situation of use.

\subsubsection{Practical Challenges with the Interaction Taxonomy in Evaluative Research}

When preparing this study and considering different technologies and
interactions, we occasionally had difficulties with Reeves and colleagues'
taxonomy~\citep{Reeves2005} for our evaluative purpose, given the multilayered
and dynamic interactions we found in real-world social situations. For many
interactions, such as phone calls or gesture-based fitness games, we could not
differentiate the manipulations and effects as clearly as initially assumed
using the four categories. Reeves and colleagues have also considered in-between
categories, for example with partially revealed or amplified effects, and from
our understanding positioned the taxonomy primarily as a useful tool for the
design phase. However, we see some challenges when using it to guide empirical
evaluation.

Take a phone call as an example. Dialing and holding the phone to one's ear are
clearly ``manipulations''. But the case is less clear for the conversation
itself. If the phone call comprises a negotiation with a call center agent about
forwarding your call, we would consider this a ``manipulative'' aspect of the
overall interaction. In contrast, laughing at a conversation partner's joke is
clearly an ``effect''.  Another example are gesture-based fitness games, where
we use body movements as input (``manipulations''). But they are also the
desired effect (i.e., to become more active). Many real-life activities have
multiple manipulations and effects, including direct interactions with the
device and gestures or mimics of the user. In our own study, the direct
``effects'' of the hearing aid (better hearing) were hidden, but of course the
``effect'' that the conversation can continue was apparent in all conditions. It
is not clear how ``no visible effect'' might have looked like in this case, and
such an isolated and binary perspective does not seem useful here.

One reason for these difficulties may be that the interaction taxonomy implies
a particular understanding of an ``interaction'', where manipulations and
effects can be clearly separated. However, other approaches exist. Hornbæk and
Oulasvirta~\citep{Hornbaek2017} identified seven different understandings of
``interaction'' in the HCI literature. The interaction taxonomy is for example
compatible with understanding interactions as a cyclic ``dialogue'' between tool
and user. The user provides some input to the system, which returns some output
that the user can sense and process, before again providing new input and so on.
In contrast, when framing interaction as experience or as embodied action, this
clear separation between user and tool, and the seemingly dyadic shift between
manipulations and effects, becomes blurry and less central. Instead, the focus
of interest shifts to questions about how the interaction in the specific
situation affects our experiences as users and as ``other people''.

In sum, from our perspective the taxonomy unfolds its main strength during the
design process. It helps designers consider what parts of an interaction to hide
or reveal, and in what specific form. However, using it in evaluative research
about real-life interactions can be challenging, because they often contain both
hidden and perceptible manipulations and effects, and in some cases these cannot
clearly be separated.

\subsubsection{Overcoming Problems of Comparing Social Situations Using Static Categories}

In the introduction, we have indicated that people are often aware of
situational differences and adapt their behavior, for example when calling on
a shopping street but not in a library. But so far, such situational differences
have not been explored systematically in the social acceptability literature.
Some studies have taken a first step using the ``Audience-and-Location Axes''
(ALA)~\citep{Rico2010} as a way to describe differences between situations. The
ALA list a handful of locations (home, pavement, driving, public transport,
pub/restaurant, workplace) and audiences (alone, partner, friends, colleagues,
strangers, family). In some studies, participants were asked to rate how
acceptable an interaction is from their perspective in each of these locations
and in front of these audiences (e.g.,~\citep{Ahlstrom2014, Alallah2018,
Rico2010}).

However, when taking an interactional approach, we see several challenges with
the ALA in their current form and use. First, the categories obviously do not
represent all social situations, and that was also not the intention when they
were first introduced~\citep{Rico2010}. Some researchers have adjusted the ALA
for their studies by adding or removing categories~\citep{Koelle2020}, but the
choice of categories nevertheless remains selective. Second, and relatedly, we
can reasonably assume experiential differences between different ``instances''
of each category. The pub/restaurant category from the ALA covers everything
from a fast food franchise to three-star restaurants, and from a shady pub to
a fancy cocktail bar, across all cultural settings.  ``Family'' includes
everyone from the toddler to the grandmother, and interpersonal relationships
between different family members may vary. Third, the ALA do not capture
overlaps between categories. A husband or a wife is both a ``partner'' and
``family'', and in some cases even a ``colleague''. A board restaurant is part
train and part restaurant. Fourth, it is not clear how much of the findings we
can transfer from one category to another. If an interaction seems unacceptable
in a restaurant but acceptable on a train, which of these rules apply in a board
restaurant? How much of this transfers to a classroom or a face-to-face
conversation? And fifth, it seems hard to communicate differences between these
categories to the HCI and design communities in detail. Our impression is that
the categories have been adopted as a checklist to recommend forms of
interaction that supposedly work ``everywhere''. The downside of this is that we
overlook more fine-grained, situation-specific solutions, which could lead to
something like a spectrum of possible forms rather than uniform,
one-form-fits-all solutions.

The interactional approach offers an alternative~\citep{Dourish2004, Uhde2021b}.
Its central claim is that we can consider people's activities as the
constitutive elements of social situations and as central for their situated
experiences. Thus, studying the relationships between these situated activities
could be a promising path forward (e.g,~\citep{Uhde2021b, Uhde2022b}). For
example, we can use typical situated activities to describe places (e.g., eating
in a restaurant and reading on a train) and interpersonal relationships (e.g.,
chatting with friends, presenting slides to colleagues). Here, the focus is on
the emerging \emph{relationships} between people and devices realized through
activities, not on location, technology, and people themselves. We would assume
that compatibility between existing, situated activities and newly introduced
interactions leads to higher social acceptability, as supported by the present
study.

By using activities as a basis, we can describe differences and similarities
between situations. For example, ``eating'' is common to all restaurants, but
``ordering at the counter'' is only common in some (e.g., food courts) and not
others (e.g., three-star restaurants). We can also represent overlapping
categories, such as a board restaurant, through the specific activities they
borrow, in this case for example eating (restaurant) and interacting with the
conductor (train). This may allow us to transfer findings from previously
studied situations to others.  So far, concrete tools to represent social
situations based on their activities (e.g.,~\citep{Uhde2022b}) are still rare.
But future work could focus on identifying common patterns of co-located
activities or other ways to describe structure in their relationships.

\subsubsection{Distinguishing Acceptable Activities From Acceptable Forms}

So far we have focused on possible acceptability differences between different
situations. But differences can also exist within the same situation, even with
similar forms of interaction. For example, ``taking a selfie'' and ``taking
a photo of a stranger'' can have quite similar forms, but they may be
experienced quite differently. Here, we want to emphasize different
considerations for the acceptability of interactions with technology that cannot
be captured when focusing primarily on form.

In other words, we want to raise the question whether a ``hidden'' or ``subtle''
form of an interaction actually makes the activity ``socially acceptable'', in
the sense that it becomes socially acknowledged as something that can or should
be performed in a given situation. This can be the case. For example, with phone
calls in public, changing aspects of its form (e.g., muting the voice of the
caller~\citep{Li2019}) has been suggested as a way to make them acceptable in
more situations (e.g., a train, a library). But in other cases, the
acceptability problem is not one of its form. Problems with phone calls during
a date may not be solved by removing ``noise'', because they rather relate to
the split attention the call implies. In such cases, the call itself is somewhat
unacceptable, independent of its form. Ens and colleagues~\citep{Ens2015} touch
upon this problem through their ``candid'' design approach that reveals certain
effects of an activity and makes it a possible topic of negotiation (e.g., ``is
browsing through social media acceptable during a work meeting?''). In some
cases, the interrelations between an activity and other co-located activities
are not clear during its introduction and have to be negotiated over time.

Regarding our own study, we would argue that there is nothing inherently wrong
about quickly readjusting a hearing aid on the level of the activity itself,
because it can be seen as necessary for the conversation. Considering the form
of interaction, we could try to make it more ``acceptable'' by completely hiding
it.  But we know from previous work that the opinions on ``visibility'' of
hearing aids vary within the community of people with divergent hearing. Some
people prefer hidden forms, and previous work suggests that this may relate to
undesirable stereotypes associated with such technologies~\citep{Schwind2019,
Schwind2020}. But other hearing aid users are quite explicit about communicating
their divergent hearing to conversation partners. They see it as a self-defining
part of themselves, and want to communicate it openly to normalize it and reduce
stigma in the society (e.g.~\citep{Doerrenbaecher2019}).

In sum, social acceptability often goes deeper than the form itself, and we
should focus on the underlying implications more explicitly in future research.

\section{Conclusion}

So far, research on social acceptability widely recommended using subtle forms
of interaction, because they seem unobtrusive and preserve
privacy~\citep{Koelle2020, Rico2010}. This makes sense in many social
situations, for example when performing technology-mediated activities among
anonymous strangers. However, in closer and more direct social interactions, or
in mixed settings, other aspects can become more important, such as paying
attention to the conversation partner~\citep{Kadylak2018, MillerOtt2015,
MillerOtt2017}. This may be better communicated through an openly visible,
``suspenseful'' hand gesture, instead of following the notion of
unobtrusiveness. To go beyond overgeneralized recommendations, we must reflect
on the activity itself and the relationships it implies between user,
``observers'' and the device. After all, ``social acceptability'' is not static:
By choosing to communicate the interaction openly through a particular form, we
may actually contribute to some desirable social change and reduce stigma---at
least in the present case of a hearing aid.


\begin{acks}

This project is funded by the \grantsponsor{501100001659}{Deutsche
  Forschungsgemeinschaft (DFG, German Research
  Foundation)}{https://doi.org/10.13039/501100001659} -- Grant
  No.~\grantnum{425827565}{425827565} and is part
  of~\grantnum{427133456}{Priority Program SPP2199 Scalable Interaction
  Paradigms for Pervasive Computing Environments}. Parts of the research
  including the prototype were funded by the German Federal Ministry of
  Education and Research (Grant No.~\grantnum{16SV7786}{16SV7786}), through the
  mEEGaHStim project.

\end{acks}

\bibliographystyle{ACM-Reference-Format}
\bibliography{bibliography.bib}


\end{document}